\documentclass[10pt, a4paper]{IEEEtran}

 \usepackage{amsmath,amssymb,amsthm}
 \usepackage{algorithmic}
 \usepackage{array}
 \usepackage{fixltx2e}
 \usepackage{stfloats}
 \usepackage[utf8]{inputenc}	
 \usepackage[british]{babel}
 \usepackage{csquotes}
 \usepackage[pdftex]{graphicx}
 \graphicspath{ {image/} }
 \usepackage{amssymb}
 \usepackage{mathtools}
  \usepackage[dvipsnames]{xcolor}

\begin{document}
\title{Time Reversal for 6G Spatiotemporal Focusing: Recent Experiments, Opportunities, and Challenges}

\author{George C. Alexandropoulos,~\IEEEmembership{Senior~Member,~IEEE,}  Ali Mokh,~\IEEEmembership{Member,~IEEE,} Ramin Khayatzadeh,\\ Julien de Rosny, Mohamed Kamoun,~\IEEEmembership{Member,~IEEE,} Abdelwaheb Ourir, Arnaud Tourin, Mathias Fink, \\and M\'{e}rouane Debbah,~\IEEEmembership{Fellow,~IEEE}
	\thanks{
		G. C. Alexandropoulos is with the Department of Informatics and Telecommunications, National and Kapodistrian University of Athens, Panepistimiopolis Ilissia, 15784 Athens, Greece and with the Technology Innovation Institute, 9639 Masdar City, Abu Dhabi, United Arab Emirates (e-mail: alexandg@di.uoa.gr).
	}	
	\thanks{
		A. Mokh, J. de Rosny, A. Ourir, A. Tourin, and M. Fink, are with ESPCI Paris, PSL Research University, CNRS, Institut Langevin, 75005 Paris, France (e-mails: \{ali.mokh, julien.derosny, a.ourir, mathias.fink\}@espci.fr, arnaud.tourin@espci.psl.eu). 	
	}
	\thanks{
	 R. Khayatzadeh and M. Kamoun are with the Mathematical and Algorithmic Sciences Lab, Paris Research Center, Huawei Technologies France, 92100 Boulogne-Billancourt, France (e-mails: \{ramin.khayatzadeh1, mohamed.kamoun\}@huawei.com).
	} 			
    \thanks{M. Debbah is with both the Technology Innovation Institute and the Mohamed Bin Zayed University of Artificial Intelligence, 9639 Masdar City, Abu Dhabi, United Arab Emirates (email: merouane.debbah@tii.ae).}
}

\maketitle

\begin{abstract}
Late visions and trends for the future sixth Generation (6G) of wireless communications advocate, among other technologies, towards the deployment of network nodes with extreme numbers of antennas and up to terahertz frequencies, as means to enable various immersive applications. However, these technologies impose several challenges in the design of radio-frequency front-ends and beamforming architectures, as well as of ultra-wideband waveforms and computationally efficient transceiver signal processing. In this article, we revisit the Time Reversal (TR) technique, which was initially experimented in acoustics, in the context of large-bandwidth 6G wireless communications, capitalizing on its high resolution spatiotemporal focusing realized with low complexity transceivers. 
We first overview representative state-of-the-art in TR-based wireless communications, identifying the key competencies and requirements of TR for efficient operation. Recent and novel experimental setups and results for the spatiotemporal focusing capability of TR at the carrier frequencies $2.5$, $36$, and $273$ GHz are then presented, demonstrating in quantitative ways the technique's effectiveness in these very different frequency bands, as well as the roles of the available bandwidth and the number of transmit antennas. We also showcase the TR potential for realizing low complexity multi-user communications. The opportunities arising from TR-based wireless communications as well as the challenges for finding their place in 6G networks, also in conjunction with other complementary candidate technologies, are highlighted.  
\end{abstract}

\IEEEpeerreviewmaketitle
\section{Introduction}
While the major telecommunications operators are currently deploying the fifth Generation (5G) of wireless networks \cite{Shafi_5G_all} in various places around the world and 3GPP has finalized the Release $17$ \cite{5GAmericas} with enhancements on the 5G New Radio (NR), academia and industry in wireless communications are already working on the definition and identification of requirements and candidate technologies for the sixth Generation (6G) communications \cite{Samsung}. The 5G NR introduces the use cases of enhanced mobile broadband, massive machine type communications, as well as ultra-reliable and low latency communications, which focus on competing performance objectives covering a wide range of vertical applications \cite{5GAmericas}. As per the latest consensus, 6G networks will require, among other metrics, higher data rates reaching up to 1 Tbps peak values, $1000\times$ network capacity as well as $10\times$ energy and cost efficiency compared to 5G, cm-level positioning accuracy, and $10^8/{\rm km}^3$ density of wireless connections \cite{saad2019vision}. To achieve the envisioned throughput requirements enabling various immersive applications (e.g., virtual and augmented reality), spectra from higher than the 5G millimeter Wave (mmWave) band as well as the TeraHertz (THz) frequencies are highly likely to be deployed, requiring cost- and power-efficient Radio-Frequency (RF) front-ends and multi-antenna transceiver architectures, as well as ultra-wideband waveforms and computationally efficient signal processing schemes.

Time Reversal (TR) is a computationally simple signal processing technique dating back to the early $90$s \cite{Fink_TR_1992}, with its initial demonstrations being for underwater wireless communications via acoustic signals. It refers to the process of transmitting a received signal in a time reversed order, profiting from the time reversibility of the wireless channel when Time Division Duplexing (TDD) is used. Radio communications based on TR exploit rich multipath propagation to offer high resolution wave focusing both in space and in time, as firstly demonstrated inside a cavity with a $2.45$ GHz Electro-Magnetic (EM) pulse in \cite{Lerosey_EM_2004}. In \cite{Liu_SPmag_2016}, TR was considered for 5G and beyond wireless communications, where it was advocated that it has the potential for realizing the benefits of massive Multiple-Input Multiple-Output (MIMO) systems using only a single-antenna base station and simple reception processing circuitry at the mobile users. The TR-based temporal focusing and the consequent received energy improvement with the increase in the number of antennas was experimentally verified in ultra-wideband communications \cite{TR_MIMO_AWPL_2011}. Multi-user communications via TR signal processing were designed in \cite{Qiu_JSAC_2006, TR_MIMO_MA_2012}. Recently, \cite{Alexandropoulos_ICASSP} demonstrated indoor cm-level localization accuracy with TR at $3.5$ GHz using channel sounding signals with up to $600$ MHz bandwidth. In \cite{xu2017waveforming}, TR was considered as a waveforming technique that treats each multipath channel component as a virtual antenna, which can be further used for coherent combining at the receiving terminal.

In this article, motivated by the large bandwidths available at mmWave and most pronouncedly at the upcoming THz wireless communications, and the respective necessity for simple transceiver signal processing and relevant hardware, as well as the recent advances with reconfigurable metamaterials for enabling programmable EM wave propagation control \cite{Philipp2019}, we present novel experimental results showcasing the high resolution spatiotemporal focusing capability of the TR technique in a wide range of carrier frequencies with very different signal propagation characteristics. Believing that this significant property, which is realized via basic transceiver operations, can contribute to 6G wireless communications in terms of coverage extension, sensing, and localization, we investigate the key requirements for the TR efficient operation and demonstrate the role of the available bandwidth and the number of transmit antennas on its performance. We discuss applications and opportunities with TR-based wireless communications, and highlight their open challenges and research directions in conjunction with complementary technologies.

\section{Time Reversal For Wireless Communications}\label{Sec:TR_for_Comms}
According to the TR framework, radio waves can be focused onto the location of a user terminal by emitting a time reversed version of the user's transmitted signal.
For example, suppose a single-antenna user emitting in the uplink a transient wave (e.g., a chirp signal), which propagates inside a complex wireless medium that can be inhomogeneous, scattering, or reverberating \cite{Lerosey_EM_2004}. The time dependence of the emitted field's components can be probed by a single- or multi-antenna base station, and then, downconverted to baseband. This resulting digital signal can be afterwards time reversed and upconverted to the analog domain for transmission in the downlink direction. Due to the reciprocity principle in TDD wireless communications, the latter downlink transmitted wave will have the same impulse response with its uplink counterpart, and will get focused on the user position. This capability has been first exploited for underwater wireless communications with acoustic signals \cite{Fink_TR_1992}, and later considered and experimented for 5G, and beyond, single- and multi-user wireless communications \cite{Liu_SPmag_2016, TR_MIMO_MA_2012}. 

The efficient application of TR requires in principle rich scattering environments, that enable multipath wave propagation, as well as large transmission/communication bandwidths \cite{Liu_SPmag_2016}. The former results in Channel Impulse Responses (CIRs) with multiple resolvable channel taps, thus increased diversity in the time domain, while the latter enables the communication system with high resolution capability for CIR estimation. Although rich scattering happens in certain wireless environments and system setups, only very recently the 5G NR introduced bandwidth values up to $800$ MHz in the mmWave frequency band \cite{5GAmericas}. We believe that these facts have prevented TR from finding its way in wireless standards up to date, however, recent trends for 6G wireless communications seem to deal with both of them, as will be discussed in the sequel.

In frequency-flat fading channels, the TR technique, that is commonly implemented in the time domain, is equivalent to Maximum Ratio Transmission (MRT), which is a spatial precoding scheme. In multi-carrier multi-antenna wireless communications with Orthogonal Frequency-Division Multiplexing (OFDM), the frequency-selective wideband channel is transformed to multiple narrowband sub-channels (or resource blocks) centered around orthogonal frequencies. In this case, MRT precoding using the conjugate frequency response for each subcarrier can be applied, which is again equivalent to performing TR precoding in the time domain. 
It is noted that MRT requires channel estimation via pilot signals for all subcarriers (this is a computationally demanding process), whereas TR is based on the CIR recording/estimation, which can be, in general, simpler. For large-bandwidth communications, such as in mmWave and for the envisioned THz in 6G, the implementation of OFDM will require increased computational complexity since significantly increased numbers of subcarriers will be utilized. 
However, with TR, the ultra-wideband channels can be estimated using an easily generated chirp signal (e.g., transmitted from the intended user in the uplink), and the TR-based precoded information-bearing signal (e.g., transmitted by a single- or multi-antenna base station in the downlink) can be realized via reversing in time the received chirp signal. This signal will be spatiotemporally focused to the user location, and thus, the user can deploy a simple single-tap receiver for information decoding, or even a non-coherent receiver \cite{mokh2017time}, when combined with ON/OFF keying or pulse-position modulation schemes. 

Apart from the potential of TR for enabling large-bandwidth wireless communications with low complexity transceiver hardware components and signal processing, its inherit spatiotemporal focusing capability has found promising applications in imaging and sensing \cite{Liu_SPmag_2016}, as well as very recently in indoor user localization \cite{Alexandropoulos_ICASSP}, wireless power transfer, and physical-layer security. As previously discussed, TR can be deployed as an alternative technology to OFDM with and without massive MIMO \cite{TR_MIMO_MA_2012} or in combination with other technologies like intelligent metasurfaces \cite{Lerosey_EM_2004} that can artificially create rich scattering channels \cite{Philipp2019}, and consequently, long CIRs with multiple resolvable channel taps. The latter applications of TR are currently being considered for 6G wireless communications \cite{saad2019vision}.

\section{Experimentation and Performance Results}
In this section, we present novel experiments and results for the performance evaluation of TR centered around the three carrier frequencies $2.5$, $36$, and $273$ GHz, which exhibit substantially different wave propagation characteristics. Initially, the considered experimental setups for the latter operating frequencies are described, and then, the TR spatiotemporal focusing capability is investigated over various system settings.    

\subsection{Experimental Setups}\label{sec:Setup}
\begin{figure*}[t!]
\centering
\includegraphics[width=\linewidth]{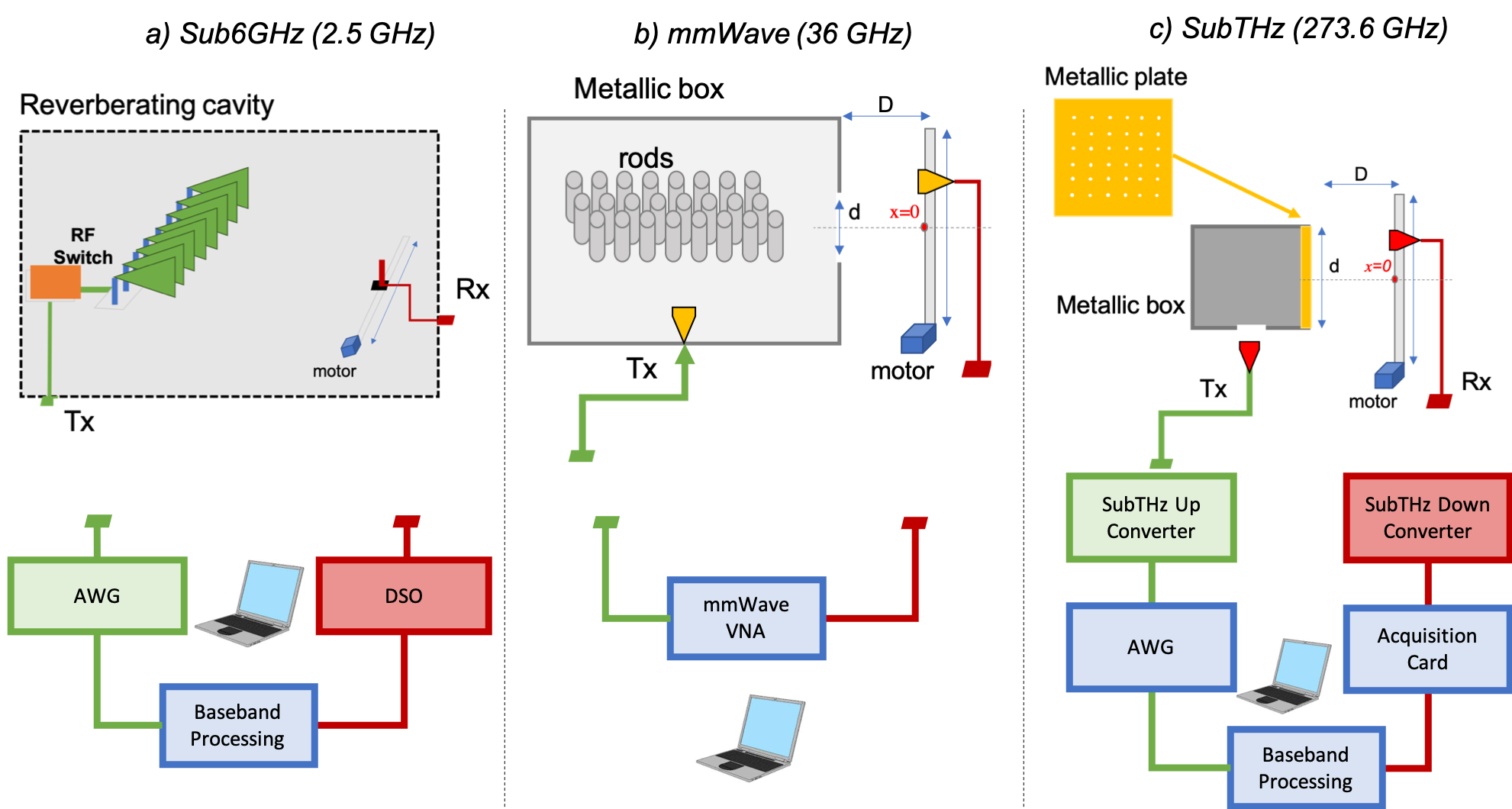}
\caption{The designed experimental setups for the investigation of the spatiotemporal focusing capability of TR-based wireless communications centered around the considered carrier frequencies: (a) $2.5$ GHz, (b) $36$ GHz, and (c) $273$ GHz. Detailed descriptions are provided in Section~\ref{sec:Setup}.}
\label{fig:setups}
\end{figure*}
Our experimental setups for realizing TR-based spatiotemporal focusing are graphically described in Fig$.$~\ref{fig:setups} for all three considered frequency bands, ranging from below $6$ GHz (sub6GHz) to subTHz (usually, between $90$ and $300$ GHz) carrier frequencies. We next describe them in detail.

\subsubsection{Sub6GHz Setup} In the experimental setup illustrated in Fig$.$~\ref{fig:setups}(a) for TR at the central frequency $f_{c}=2.5$ GHz with variable bandwidth $B$, we have considered a Tx composed of a linear array with $N_t=8$ Vivaldi antenna elements, all connected to a solid-state RF switch, whose input is fed with the amplified signal of a Tektronix AWG-7012  Arbitrary Waveform Generator (AWG), operating with a sampling rate of $10$ GS/s. A dipole antenna was used at the Rx side, which was placed on a motorized axis in order to realize variable cm-spaced reception positions. This antenna was connected to the low noise amplifier of the Rx RF chain, and then, to the Tektronix TDS6604B Digital Storage Oscilloscope (DSO) working at the symbol rate $12.5$ GS/s. The experiment took place inside a three-dimensional reverberating chamber of size 1.5 m $\times$ 1 m $\times$ 0.5 m in order to create a rich multipath channel, which is essential for realizing efficient TR. The estimation of the CIR and the TR precoding were performed, as follows. First, a chirp signal spanning the frequency range [$f_c-B/2$ $f_c+B/2$] was generated by the AWG and then transmitted by the Tx antenna. The received signal on the Rx antenna was then synchronously recorded by the DSO and stored in baseband representation into a desktop computer, which was assigned the implementation of TR processing. In particular, the computer performed CIR estimation by deconvoluting the latter baseband received signal with the transmitted chirp signal. This CIR estimation was flipped in the time domain and then resampled from $12.5$ to $10$ GS/s, as needed by the deployed AWG.

\subsubsection{mmWave Setup} 
The realized mmWave setup at the central frequency $f_{c}=36$ GHz with bandwidth $B=2$ GHz deploys a 30 cm $\times$ 20 cm $\times$ 20 cm closed metallic box, as depicted in Fig$.$~\ref{fig:setups}(b), that includes multiple closely placed vertical rods in order to enable multiple reflections of the transmitted signal. As shown in the figure, the side of the box that faces the Rx antenna has a square window of width $d=5$ cm, while a similar Tx antenna is placed on another side of the box that is perpendicular to the window. We have used A-info Octave horn antennas for both the Tx and Rx, having wideband operation from 18 to 40 GHz with 20 dB gain. The orientation of the Tx antenna was such that the emitted waveform faces the rods before reaching the window, and consequently, the Rx antenna. The latter antenna was placed on a motorized axis at a distance $D=20$ cm from the box, enabling its parallel movement to the box's window. We considered that the point $x=0$ on the axis, where the Rx antenna lies, faces the center of the metallic box. 

Instead of using an AWG and a DSO in our designed mmWave experiment, as we did for the sub6GHz one, we have deployed the Vector Network Analyzer (VNA) R\&S ZNA43 to sound in the frequency domain the channel between the Tx and Rx antennas. This channel sounding was performed, as previously mentioned, around $36$ GHz with $2$ GHz of bandwidth. In order to emulate the TR processing, we first numerically transformed the channel's frequency response, which was probed by the VNA, to the CIR using the discrete Fourier transform, and then flipped it in the time domain. Finally, this time-domain signal was convoluted with all probed CIRs. 

\subsubsection{SubTHz Setup} 
Our experimental setup at $f_{c}=273.6$ GHz using a 5 cm $\times$ 5 cm $\times$ 5 cm closed metallic box is demonstrated in Fig$.$~\ref{fig:setups}(c). Similar to the mmWave case, to create rich multipath needed for efficient TR operation, we have created a small window at an one side of the box, with its perpendicular side being opened and replaced with a metallic square plate having multiple small holes with diameter less than 1 mm. Both the Tx and Rx are equipped with identical WR-$2.8$ horn antennas (operating from $260$ up to $400$ GHz), the former facing the small window, and the latter placed on a motorized axis, as shown in Fig$.$~\ref{fig:setups}(c). The axis of the Rx antenna movement is parallel to the plate, consequently to the one side of the box, with a separating distance $D=5$ cm. As shown in the figure, the point $x=0$ on the axis faces the center of the metallic box. 

To create transmitted signals at subTHz, we generated a reference sinusoidal signal at the desired frequency $f_{c}=273.6$ GHz, as follows. The deployed local oscillator R\&S SGMA produced a sinusoid at $5.7$ GHz, which was then up-converted for transmission using a WR2.2 VDI SAX WM570 harmonic mixer (operating at $260$ up to $400$ GHz with $-10$ dBm transmit power) with a multiplication factor set to $48$ (note that $5.7\times48=273.6$). At the Rx side, a harmonic mixer with the same multiplication factor was deployed for down-conversion back to $5.7$ GHz. Both the AWG to generate the baseband signal of bandwidth $3$ GHz, and the acquisition card to detect it, were connected to the same clock reference working at $10$ MHz. The CIR estimation was performed similar to the sub6GHz experiment, and TR precoding was simulated in the desktop computer similar to the mmWave experiment.

\subsection{Results for Sub6GHz}
\begin{figure}[t!]
\centering
\includegraphics[width=\columnwidth]{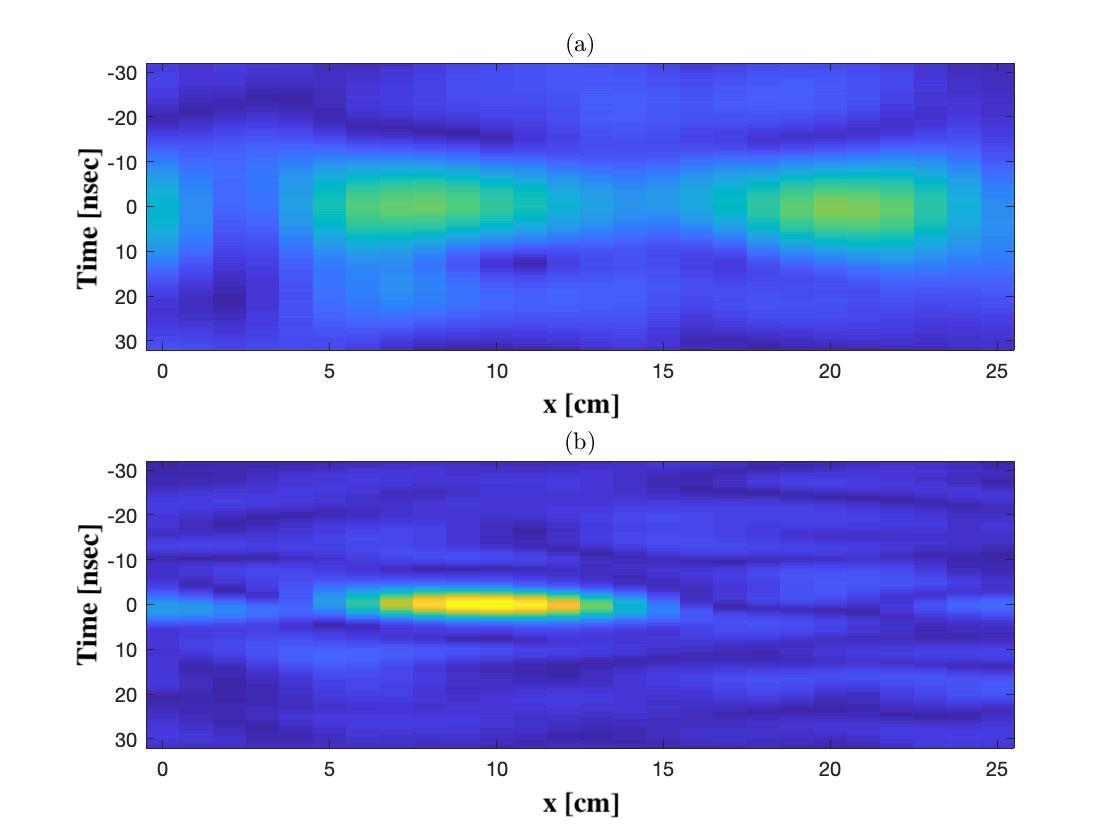}
\caption{Measurements for the TR spatiotemporal focusing capability of the $8\times1$ MISO wireless system at different Rx positions and different bandwidth values $B$: (a) two Rx positions at $x=7.5$ cm and $x=20$ cm with $B=100$ MHz; and (b) the Rx is positioned at $x=10$ cm and $B=400$ MHz.  
}\label{3Pos}
\end{figure}

The impact of the number of the transmit antennas $N_t$ and the communication bandwidth $B$ in the spatiotemporal focusing capability of TR precoding, as implemented in the experimental setup of Fig$.$~\ref{fig:setups}(a), 
was thoroughly investigated and representative results are illustrated in Fig$.$~\ref{3Pos}. In particular, we have first emulated a TR-processed CIR of power $-4$ dBm and with bandwidth values $B=\{100,200,400\}$ MHz to focus towards the position $x=10$ cm, and measured the baseband received signal strength at $30$ distinct Rx positions, ranging from the point $x=0$ cm in the motorized axis to $x=30$ cm, with a space separation of $1$ cm. When $B=100$ MHz was considered for $N_t=1$ (a Single Input and Single Output (SISO) system) and for two MISO configurations for $N_t=2$ and $8$, the received signal strength indeed took its maximum value around $x=10$ cm, while it was shown that this value increases with increasing $N_t$. More specifically, for the $8\times1$ MISO wireless system that yielded the best TR-based spatiotemporal focusing capability, it was corroborated that the received signal strength gets its maximum value around the intended Rx position with a spatial resolution of $6$ cm and in $10$ nsec temporal resolution. It is noted that the resulted spatial width of the signal focusing area is approximately equal to the half of the signal wavelength; recall that $f_{c}=2.5$ GHz.
 
The potential of TR precoding to create less interference in unintended Rx positions, thanks to its spatiotemporal focusing efficiency, and the role of the bandwidth $B$ on its time focusing capability are demonstrated in Fig$.$~\ref{3Pos} for the considered $8\times1$ MISO wireless system. 
As depicted in Fig$.$~\ref{3Pos}(a) for $B=100$ MHz, the spatial resolution of $6$ cm is sufficient to focus the transmitted signal simultaneously towards the two different positions at $x=7.5$ cm and $x=20$ cm. Considering the position $x=10$ cm and $B=400$ MHz, Fig$.$~\ref{3Pos}(b) showcases that increasing $B$ results in improved temporal focusing of the received signal. It is particularly depicted in this subfigure that the time resolution of the received signal is approximately equal to $2.5$ nsec, while for $B=100$ and $200$ MHz the resolution was $10$ and $5$ nsec, respectively. In conclusion, our experimental results prove that the TR-based time resolution is approximately equal to the inverse of the signal bandwidth.

\subsection{Results for mmWave}\label{Sec:mmWave}
\begin{figure}
\centering
\includegraphics[width=\columnwidth]{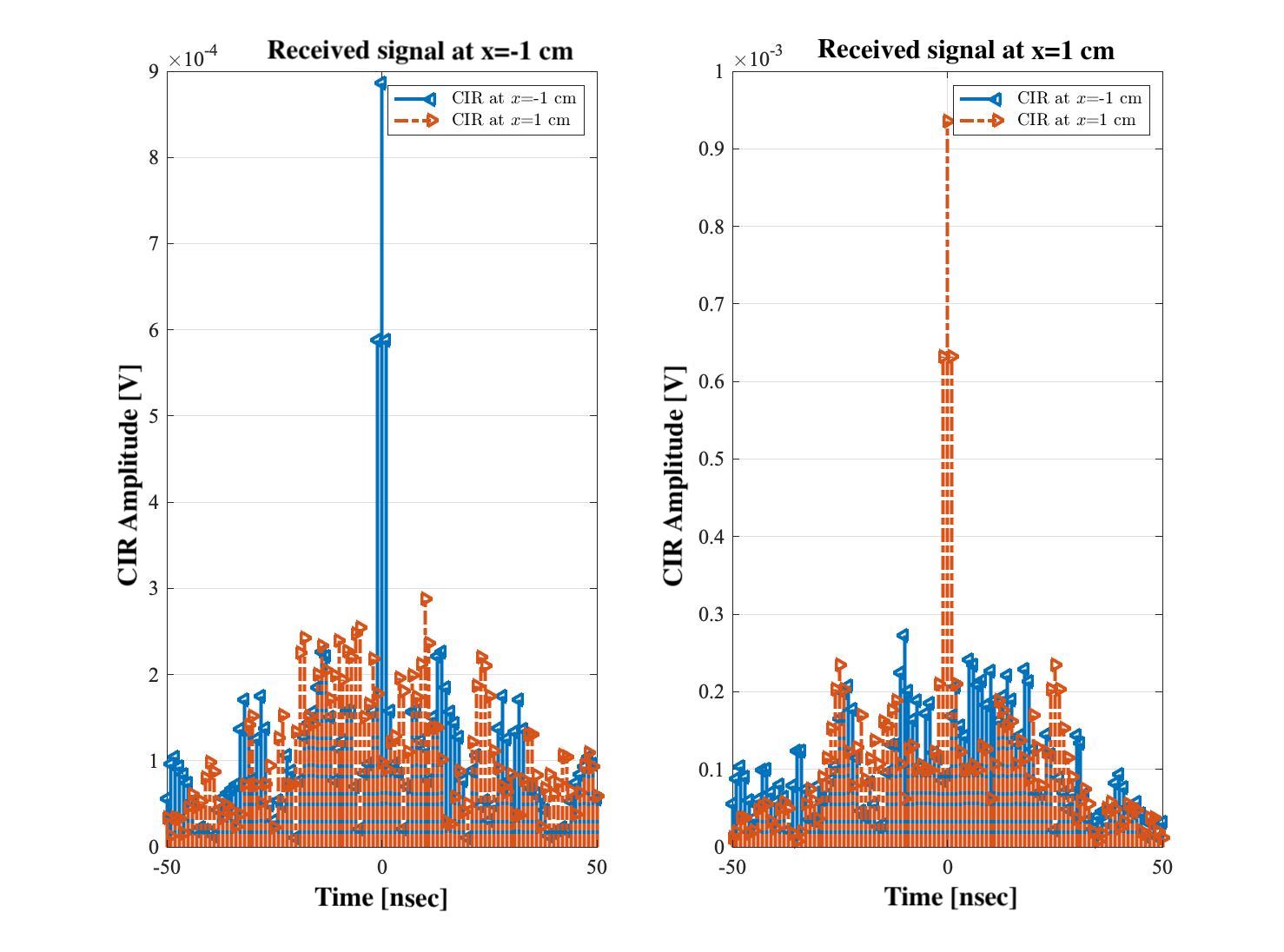}
\caption{The amplitudes in Volts of the CIR components for the mmWave system in Fig$.$~\ref{fig:setups}(b) operating at $36$ GHz with bandwidth $B=2$ GHz and considering two Rx positions at the points $x=-1$ cm and $x=1$ cm. TR is applied for the former Rx position on the left, while for the latter position on the right.} \label{fig:mmWave}
\end{figure}

The capability of TR to spatiotemporally separate the two Rx positions  at $x=-1$ cm and $x=1$ cm is illustrated in Fig$.$~\ref{fig:mmWave}. In particular, the amplitudes in Volts of the CIR components when TR is applied at any of the two positions is depicted. It is evident that by applying TR, we are capable to focus two different signals towards the two intended Rx positions. In both subfigures, when targeting a specific position, a focusing signal at the time instant $t$=0 is significantly higher than the level of interference (inter-user and inter-symbol interference), which implies that space multiplexing can be achieved thanks to TR in this frequency range. This means that with TR, we can send two different data streams towards two different Rx positions separated by $2$ cm; this has been termed in past theoretical works as the TR division multiple access (TRDMA) scheme \cite{TR_MIMO_MA_2012}.

\subsection{Results for SubTHz}
In Fig$.$~\ref{thzsetp}, the TR spatiotemporal focusing capability with the subTHz experimental setup of Fig$.$~\ref{fig:setups}(c) is illustrated. We have transmitted a signal of power $-10$ dBm covering a bandwidth $B=3$ GHz, and measured the received signal strength at $21$ distinct positions of the Rx antenna in the motorized axis, ranging from the point $x=-3$ mm to $x=3$ mm, with a space resolution of $0.3$ mm. As shown in Figs.~\ref{thzsetp}(b) and~\ref{thzsetp}(c), the width of the TR spatial focusing capability (i.e., the yellow colored area) in both considered time-orthogonal true Rx positions is approximately equal to the signal wavelength, which is $1$ mm, and the temporal resolution is less than $1$ nsec. It is also depicted in Fig.~\ref{thzsetp}(a) for the designed experimental setup that, no spatiotemporal focusing at the intended Rx positions is feasible when TR processing is not applied.    

\section{Opportunities and Open Challenges}
The experimental results of the previous section for three carrier frequencies with very different signal propagation characteristics and large, but reasonably large for beyond 5G, transmission bandwidths showcased that TR can offer signal spatial focusing approximately equal to the wavelength with temporal resolution of the order of the inverse of the transmission bandwidth, i.e., of few nanoseconds. In this section, we discuss representative opportunities for future wireless communications resulting from TR's capabilities, while identifying key open challenges for TR to be a strong candidate technology for 6G wireless communications.

\subsection{Opportunities}
As discussed in Section~\ref{Sec:TR_for_Comms}, the TR technique requires the estimation of the CIR, which can be realized with simple signal processing from the Tx that generates the information data. Then, the Rx can deploy a  simple reception module for information decoding. To this end, in TDD communications, the CIR can be estimated in a dedicated uplink phase followed by the TR-precoded information transmission in the downlink. We next present some of the key opportunities with TR:
\begin{figure}
\centering
\includegraphics[width=\columnwidth]{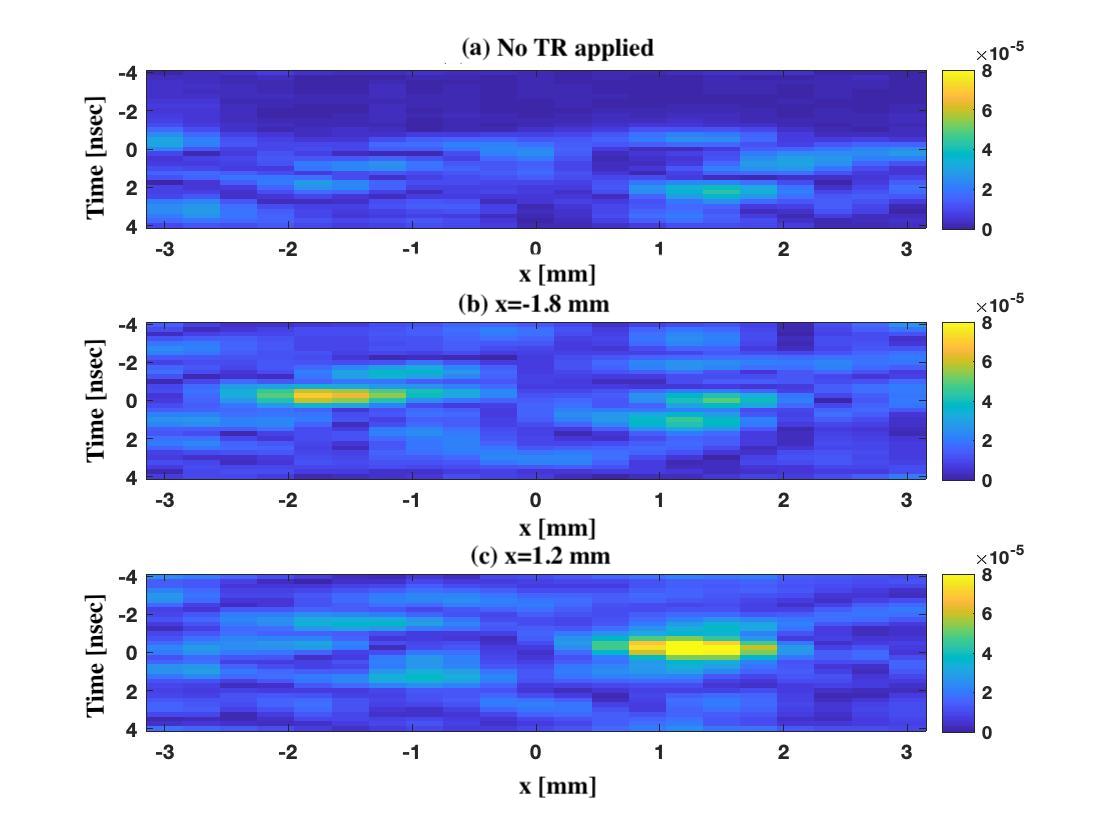}
\caption{The TR spatiotemporal focusing capability of the subTHz system in Fig$.$~\ref{fig:setups}(c) operating at $273.6$ GHz with bandwidth $B=3$ GHz for different Rx positions. The received signal strength without applying TR is also sketched.}
\label{thzsetp}
\end{figure}

\textbf{Inherent Localization and Sensing:} The high resolution spatiotemporal focusing capability of the TR technique, as showcased in all experimental results in Section~\ref{Sec:TR_for_Comms} and in the recent laboratory experiments in \cite{Alexandropoulos_ICASSP}, can be naturally exploited for localization and sensing. The latter two applications are currently highly relevant for 6G wireless communications \cite{saad2019vision}, both as end services as well as prerequisites for several high performance applications (e.g., immersive and vehicular communications, and integrated communications and radar). As it has been demonstrated in the available results for indoor environments, TR precoding offers signal focusing at an intended Rx position with cm-level accuracy. This feature enables highly localized signal delivery, increasing signal coverage and reducing signal leakage (i.e., interference) at nearby Rx positions. The CIR estimation, which is the initial phase of the TR technique, can be also incorporated with machine learning techniques to map the CIR referring to an intended Rx position to absolute location estimation. Hence, TR can be considered either alone or synergistically  with other super-resolution approaches for high precision positioning systems.

High resolution imaging, which encompasses the family of approaches for sensing cooperating active users in a given wireless environment surveilled by a plurality of transceivers, has also relied on variants of the TR technique. Combined with super-resolution signal classification techniques, TR has been applied for environmental imaging of point targets from arbitrary sensor array geometries, with applications in seismo- and ocean-acoustics, and ultrasonic nondestructive evaluation systems. This TR-enabled feature can be exploited for designing radio cartography maps that can be used for spatiotemporal resource allocation optimization and service provisioning.

\textbf{Low Complexity Reception:} In contrast to OFDM that requires increased computational complexity which increases significantly with the number of subcarriers, TR relies on time-domain channel estimation (i.e., CIR estimation), using for example chirp signals, and enables simple, even single-tap, reception. This fact can motivate the adoption of low complexity TR-based receivers for user equipment in large-bandwidth wireless communications, such as in mmWave and for the provisioned THz bands in 6G. In addition, TR can be further combined with spatial modulation for facilitating low complexity reception for single-carrier wireless communications, enabling non-coherent signal detection (e.g., via ON/OFF keying or pulse-position modulation) at battery-limited and computationally-basic devices intended for the Internet of Things (IoT). For example, in the recent work \cite{phan2019single}, TR was successfully deployed for receive spatial modulation, where IoT devices were equipped with one reconfigurable antenna and a non-coherent signal detector.   

\textbf{Multi-User Transmissions:} The interpretation of the CIR results in Fig$.$~\ref{fig:mmWave} corroborated the potential of TR for multi-user spatial multiplexing. It was specifically advocated that TR can be used to concurrently transmit two different data streams towards two cm-separated Rx positions. There has also been increased interest in the relevant literature for efficient signal processing techniques for reducing the effect of the unavoidable interference caused by conventional TR precoding. To this end, an iterative TR process requiring feedback between the Tx and a given Rx has been introduced, according to which, the interference is iteratively subtracted at that Rx using a precoded TR signal. The concept of TRDMA \cite{TR_MIMO_MA_2012} has been also theoretically shown to provide a cost-effective waveform for multiple access of energy-efficient IoT nodes. Hence, TR offers various perspectives for large numbers of simultaneous transmissions intended for multiple low complexity receivers.

\subsection{Research Challenges}
The designed experimental setups for TR in Fig.~\ref{fig:setups} include either a reverberating cavity or metallic boxes that mechanically create strong multipath signal propagation. Similar rich scattering conditions were created for \cite{Alexandropoulos_ICASSP}'s laboratory experiments, where it was found that, for channels centered at $3.5$ GHz with at least $10$ strong multipath components and more than $100$ MHz channel sounding capability, TR-based precoding provides less than $5$ cm spatiotemporal focusing and up to $4$ dB signal coverage improvement. Given the requirements for efficient TR operation, we next discuss some of the key challenges with TR and its emerging applications:

\textbf{EM Wave Propagation Control:} The efficacy of TR highly depends on the diversity in signal propagation paths in time. The more the number of paths, the better is the performance of TR. In fact, for scenarios with poor multipath, TR is usually inefficient. This means that TR should be applied only in physically or artificially-customized rich scattering scenarios. For the latter case, intelligent reflecting metasurfaces \cite{Philipp2019} can be deployed either close to the Tx, or Rx, or both, or even being a part of each of them (depending on the carrier frequency), whose operation needs to be jointly optimized with TR precoding. To this end, the interplay between multipath richness and TR efficiency needs to be fully characterized in order to be dynamically manipulated via low-overhead algorithms by reconfigurable metasurfaces.   

\textbf{Interference Management:} As depicted in Fig$.$~\ref{fig:mmWave}, TR focuses the signal in time and space by maximizing the central peak at the intended Rx position. However, this approach does not annihilate interference neither in time nor in space, possibly resulting in inter-symbol and inter-user interference that can limit the performance of TR-based multi-user wireless communication systems. This problem is more severe when there exist spatially correlated signal propagation paths among the plurality of the users. To alleviate such interference problems, efficient interference-handling techniques for TR need to be designed.

\textbf{Mobility and Synchronization:} The TR technique relies on the reciprocity property of wireless communication channels. If the Tx and/or the Rx are mobile, the received signal level varies with the node's speed. If that speed is higher than what the channel coherence time can handle, the CIR estimation could be outdated, resulting in spatiotemporal focusing for an erroneous Rx position. Hence, efficient synchronization methods are necessary for TR-linked nodes that dynamically capture the spatiotemporal coherence of the wireless channel.

\textbf{Fragmented Large Bandwidth:} As previously discussed, the availability of large communication bandwidths enables high resolution CIR estimation, which is essential for TR. However, in sub6GHz applications, large communication bandwidths are either infeasible or can only be available via carrier aggregation. It is appealing to investigate the design of TR techniques for large-bandwidth communications resulting from the aggregation of fragmented bandwidth chunks. 

\textbf{Synergies with Other Technologies:} Despite the promising experimental results with TR, advances in algorithmic design and experimentation for TR-precoded data communication considering single-user multi-stream MIMO and multi-user massive MIMO systems are required. In addition, the adequate regimes and applications of TR need to be fully characterized in conjunction with the technique's promising synergies with other candidate 6G technologies, like holographic MIMO and orbital angular momentum, which exhibit potential for enabling service delivery with high spatiotemporal resolution and multi-user multiplexing. Finally, the interplay of TR with physical-layer security necessitates investigation to unveil the TR potential for CIR-based information encryption.

\section{Conclusion}
In this paper, inspired by the current trends in mmWave and THz technologies as well as in localization and sensing for beyond 5G networks, we have experimentally investigated the spatiotemporal focusing capability of the TR technique in large-bandwidth wireless communications with various carrier frequencies, ranging from sub6GHz to subTHz. We discussed the underlying principle of TR with some representative demonstrations and considerations up to date, and presented its key competencies in terms of low complexity transceiver hardware and signal processing, as well as its requirements for rich scattering and large communication bandwidth. Our experimental results showcased the roles of the frequency band and its width, as well as of the number of transmit antennas, in the TR spatiotemporal focusing in a single and multiple spots. Indicatively, it was demonstrated for the first time in the subTHz domain that TR is capable of offering spatial focusing of less than $1$ mm, and less than $1$ nsec temporal resolution with a $3$ GHz bandwidth. We finally highlighted the opportunities and challenges arising from TR in 6G wireless communications and discussed its promising synergies with other emerging technologies. 
\bibliographystyle{IEEEtran}
\bibliography{IEEEabrv,ref}
\end{document}